%% file: standard_eccentricity.tex
\documentclass[twocolumn]{aastex631}
\usepackage[T1]{fontenc}
\usepackage[utf8]{inputenc}
\usepackage{lmodern}
\usepackage{verbatim}
\usepackage{physics}
\usepackage{acronym}
\usepackage{amsmath}
\usepackage{orcidlink}
\graphicspath{{figures/}}
\usepackage{hyperref}

\newcommand{\corrections}[1]{{{#1}}}

\newcommand{\ICTS}{\affiliation{International Centre for Theoretical Sciences, Tata Institute of Fundamental Research, Bangalore  560089, India}}
\newcommand{\CITA}{\affiliation{Canadian Institute for Theoretical Astrophysics, University of Toronto, 60 St George St,  Toronto, ON M5S 3H8, Canada}}
\newcommand{\UChicago}{\affiliation{Department of Physics, The University of Chicago, 5640 South Ellis Avenue, Chicago, Illinois 60637, USA}}
\newcommand{\KICP}{\affiliation{Kavli Institute for Cosmological Physics, The University of Chicago, 5640 South Ellis Avenue, Chicago, Illinois 60637, USA}}
\newcommand{\EFI}{\affiliation{Enrico Fermi Institute, The University of Chicago, 933 East 56th Street, Chicago, Illinois 60637, USA}}
\newcommand{\Adler}{\affiliation{Adler Planetarium, 1300 South DuSable Lake Shore Drive, Chicago, IL, 60605, USA}}
\newcommand{\CIERA}{\affiliation{Center for Interdisciplinary Exploration and Research in Astrophysics (CIERA), Northwestern University, Evanston, IL, 60201, USA}}

\usepackage{appendix}

\begin{document}

\title{Consistent eccentricities for gravitational wave astronomy: Resolving discrepancies between astrophysical simulations and waveform models}

\author[0000-0002-4103-0666]{Aditya Vijaykumar}
\email{aditya@utoronto.ca}
\CITA
\ICTS

\author[0000-0002-8304-0109]{Alexandra G. Hanselman}
\UChicago

\author[0000-0002-0147-0835]{Michael Zevin}
\KICP
\EFI
\Adler
\CIERA

\begin{abstract}
Detecting imprints of orbital eccentricity in \ac{GW} signals promises to shed light on the formation mechanisms of binary black holes. To constrain formation mechanisms, distributions of eccentricity derived from numerical simulations of astrophysical formation channels are compared to the estimates of eccentricity inferred from \ac{GW} signals. We report that the definition of eccentricity typically used in astrophysical simulations is inconsistent with the one used while modeling \ac{GW} signals, with the differences mainly arising due to the choice of reference frequency used in both cases. We also posit a prescription for calculating eccentricity from astrophysical simulations by evolving ordinary differential equations obtained from post-Newtonian theory, and using the dominant ($\ell = m =2$) mode's frequency as the reference frequency; this ensures consistency in the definitions.
On comparing the existing eccentricities of binaries present in the Cluster Monte Carlo catalog of globular cluster simulations with the eccentricities calculated using the prescription presented here, we find a significant discrepancy at $e \gtrsim 0.2$; this discrepancy becomes worse with increasing eccentricity. We note the implications this discrepancy has for existing studies, and recommend that care be taken when comparing data-driven constraints on eccentricity to expectations from astrophysical formation channels.
\end{abstract}

\input{commands}
\input{values}
\input{acronyms}

\section{Introduction}

Definitive measurement of orbital eccentricity in a \ac{CBC} signal 
is one of the most exciting prospects of \acf{GW} astrophysics. 
Despite $\mathcal{O}(100)$ \ac{BBH} signals being detected by the \ac{LVK} network~\citep{2015CQGra..32g4001L,2015CQGra..32b4001A,2013PhRvD..88d3007A,2021PTEP.2021eA101A,2018LRR....21....3A} to date~\citep{2023PhRvX..13d1039A}, the distinct pathways that led to the formation and merger of these \ac{BBH} signals remains a mystery. 
Measurements of masses, spins, and redshift evolution have helped to identify broad constraints on the underlying physical processes that lead to \ac{BBH} formation~\citep{2023PhRvX..13a1048A}, although measurement uncertainties paired with the inherent uncertainties embedded within formation channel modeling make it difficult to use these measurables to robustly pin down the formation pathway for a single \ac{BBH} or the relative fraction of \acp{BBH} that results from one formation channel compared to another (see e.g.~\citealt{2021ApJ...910..152Z,2023ApJ...955..127C}). 

Measurable orbital eccentricity in stellar-mass \ac{BBH} mergers, on the other hand, has been shown to be a robust indicator that a particular \ac{BBH} signal resulted from a small subset of viable channels~\citep{2009MNRAS.395.2127O,
2012PhRvD..85l3005K,
2014ApJ...784...71S,
2017ApJ...841...77A,
2017ApJ...840L..14S,
2017ApJ...836...39S,
2018PhRvD..98l3005R,
2018ApJ...863....7R,
2018PhRvD..97j3014S,
2018ApJ...855..124S,
2018arXiv180208654S,
2019MNRAS.486.4781F,
2019MNRAS.488.4370F,
2019ApJ...881...41L,
2019MNRAS.483.4060L,
2019ApJ...871...91Z,
2020MNRAS.498.4924M,
2021MNRAS.506.1665G,
2021ApJ...907L..20T,
2021MNRAS.505..339M,
2021ApJ...921L..43Z,
2021ApJ...923..139Z,
2022Natur.603..237S,
2022MNRAS.511.1362T,
2023arXiv230307421D,
2023PhRvD.107l2001R,
2023arXiv231213281T}. 
Eccentricity is efficiently damped as a compact binary system inspirals and loses energy and angular momentum through \ac{GW} emission~\citep{1964PhRv..136.1224P}. 
To retain any measurable semblance of orbital eccentricities at frequencies accessible to current ground-based \ac{GW} detectors, a \ac{BBH} system therefore must be able to form at a high eccentricity ($\gtrsim 0.9$) and merge on an extremely rapid timescale ($\lesssim 1~\text{yr}$, see \citealt{2019ApJ...871...91Z}) or have some mechanism that continually pumps eccentricity into an inspiraling system \corrections{(e.g., the von Zeipel-Lidov-Kozai mechanism \citealt{vonZeipel1910,LIDOV1962719,1962AJ.....67..591K, 2016ARA&A..54..441N})}. 

The former of these possibilities is something that can only be attained through strong dynamical encounters within dense stellar environments, such as globular clusters, nuclear clusters, and the disks of active galactic nuclei. 
The synthesis of eccentric \acp{BBH} mergers in globular clusters is a robust and unavoidable result of simple physical processes such as two-body relaxation and (chaotic) gravitational encounters between three or more black holes within the cluster core (see e.g.~\citealt{2018PhRvD..98l3005R}). 
Since the relative fraction of systems that proceed through this eccentric subchannel is relatively robust to uncertainties in cluster environments, the detection and measurement of even a small number of eccentric \ac{BBH} mergers can lead to stringent constraints on the relative fraction of merging \ac{BBH} systems that result from dynamical environments entirely~\citep{2021ApJ...921L..43Z}. 

Although eccentricity is a powerful constraint on formation pathways in theory, in practice the measurement and interpretation of an eccentric \ac{GW} signal is much more muddled. 
Current matched-filter \ac{CBC} searches applied to \ac{GW} data~\citep{2023PhRvX..13a1048A} only use aligned-spin and quasi-circular templates~\citep{2019arXiv190108580S,2020PhRvD.102b2004D,2021CQGra..38i5004A}, and thus a selection effect exists against detecting eccentric \ac{GW} signals. The selection function for eccentric systems due to this method for searching the data has yet to be adequately characterized (see discussion in \citealt{2021ApJ...921L..43Z}).  Waveform-independent searches for eccentric mergers have also been carried out~\citep{2019PhRvD.100f4064A,2023arXiv230803822T} with no significant triggers; this is unsurprising, since these searches are typically only sensitive to short duration (i.e. high mass) systems.
A number of eccentric waveform approximants have been built~\citep{2016PhRvD..93f4031T, 2018PhRvD..97b4031H, 2021PhRvD.103j4021N,2022PhRvD.105d4035R,2022CQGra..39c5009L}, although much more work needs to be done to make them accurate over a large range of eccentricities.  
In particular, simultaneously modeling eccentricity and spin precession has proved to be challenging (see \citealt{2023arXiv231004552L} for a recent attempt\footnote{See also \citet{2021arXiv210610291K} that constructs an approximant that includes eccentricity and precession for binaries detectable with LISA}), and these properties have degenerate effects on the \ac{GW} waveform making it difficult to distinguish between precessing and eccentric hypotheses especially for massive signals~\citep{2023MNRAS.519.5352R,2023PhRvD.107j3049X,2020ApJ...903L...5R}.
Lastly, the definition of eccentricity differs between \ac{GW} waveform models. Eccentricity evolves during the inspiral of a binary, and hence can only be defined with reference to an epoch in the orbit of that binary. 
However, the definition of this reference epoch and of eccentricity itself is inconsistent across different state-of-the-art eccentric waveform approximants, \corrections{in part because eccentricity is a gauge-dependent quantity within general relativity}. There have been attempts towards bringing all these definitions on an equal footing~\citep{2022ApJ...936..172K,2022PhRvD.105d4035R, 2023PhRvD.107f4024B, 2023PhRvD.108j4007S}, allowing consistent comparison of eccentricities derived from diverse waveform approximants. 

A robust and consistent definition for eccentricity that is shared across waveform models and astrophysical models is imperative to yield any astrophysical conclusions in the event of an eccentric \ac{BBH} detection. In this paper, we show that the definition of eccentricity currently employed in astrophysical models is different from the definition used in waveform models. \corrections{This discrepancy between these definitions is relevant for eccentricity $e \gtrsim 0.2$ with a factor of $\sim 2$ discrepancy at high $e$.}
We also paint a clearer picture as to where these discrepancies in the definition of eccentricity lie, and present a framework for determining a robust definition of eccentricity that can be utilized for self-consistent comparisons between predictions from the astrophysical modeling community and the \ac{GW} waveform community.

Our paper is structured as follows. 
In Section~\ref{sec:background}, we review current methods that are used for defining eccentricity, both in astrophysical simulations and waveform modeling. 
Section~\ref{sec:standardized_eccentricity} presents a standardized definition of eccentricity and a scheme for extracting this definition of eccentricity from astrophysical simulations for a direct comparison to \ac{GW} waveforms. 
We also show how quoted eccentricity values significantly differ when a self-consistent definition of eccentricity is not used, investigate the robustness of our improved definition to higher-order \ac{PN} effects and spin, and discuss how our self-consistent picture of eccentricity affects the inferred fraction of measurably eccentric sources from astrophysical models. 
Finally, we provide a summary and concluding remarks in Section~\ref{sec:conclusions}. 
\corrections{Example python scripts~\citep{vijaykumar_2024_10974975} and data products~\citep{vijaykumar_2024_10633005} are available on Zenodo}.

\subsection{Definitions of eccentricity used in this work}
This work refers to eccentricity in many different contexts. For ease of reading, we provide a list of these below.
\begin{enumerate}
    \item $e_{\rm W03}$: Eccentricities calculated from the prescription outlined in \citet{2003ApJ...598..419W}. This prescription has been commonly used to extract eccentricities from  astrophysical simulations of binary formation (e.g.~\citealt{2017ApJ...840L..14S,2018PhRvD..97j3014S,2018PhRvD..98l3005R,2019ApJ...871...91Z,2021ApJ...921L..43Z, 2022arXiv221010055K, 2023MNRAS.526.4908C}). As we shall see in Section~\ref{subsec:WenFormalism}, this prescription uses the \ac{GW} peak frequency as the reference frequency.
    \item $e_t^{k {\rm PN}}$: The eccentricity computed by evolving \ac{PN} equations accurate to the $k$-th order (see e.g.~\citealt{2016PhRvD..93f4031T,2018PhRvD..98j4043K}). \corrections{This uses the dominant, quadrupolar mode ($\ell = m = 2$) \ac{GW} frequency as the reference frequency}. 
    \item $e_{\rm W03}^{f_{22}}$: Eccentricities calculated using the W03 prescription, but extracted at fixed 22-mode frequency.
    \item $e_{\rm gw}^{\rm wf}$: Eccentricity definition proposed in \citet{2023PhRvD.108j4007S}, computed from a gravitational waveform using the \texttt{gw\_eccentricity} package (see also \citealt{2022PhRvD.105d4035R}).
    \item $e_0$:  Eccentricity at a given reference point from astrophysical simulations, which is used as an initial condition (along with initial separation $a_0$) to evolve the system to a reference point suitable for \ac{GW} detectors. 
\end{enumerate}

\section{Background}\label{sec:background}
\subsection{Peak harmonic of a \ac{GW} signal in presence of eccentricity}
\begin{figure}
    \centering
    \includegraphics{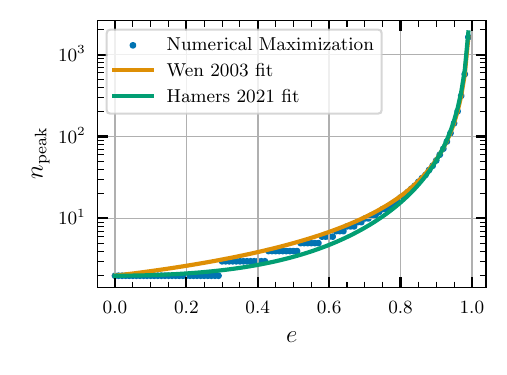}
    \caption{Numerical solution of Eq.~\eqref{eq:peak_harmonic} for the peak harmonic plotted along with the analytical fits from \citet{2003ApJ...598..419W} and \citet{2021RNAAS...5..275H}. We see that the peak harmonic increases with eccentricity, and reaches values $> 100$ for $e > 0.9$. We also see that the fit in \citet{2021RNAAS...5..275H}  fits the numerical solution better, especially at lower values of eccentricity.}
    \label{fig:peak_harmonic}
\end{figure}
At least in the inspiral phase of the binary, {i.e.} when the dynamics is well-described by \ac{PN} expansion, a binary emits \acp{GW} at harmonics of its orbital frequency $f_\mathrm{orb}$, $$f_{\mathrm{GW} , n} = n f_\mathrm{orb} \qq{;} n \geq 2 \qq{.}$$
Note that the harmonic index $n$ is different from the $(\ell, m)$ basis used to parametrize \ac{GW} polarizations. At leading order, power emitted by the $n$-th harmonic is given by~\citep{1963PhRv..131..435P},
\begin{align}
\label{eq:peak_harmonic}
P_n = \frac{32}{5} \frac{G^4}{c^5} \frac{m_1^2 m_2^2 M}{a^5} g(n,e),
\end{align}
where the function $g(n,e)$ quantifies the enhancement in the power emitted in the $n^\mathrm{th}$ harmonic compared to a circular orbit (where there is no emission at $n\neq2$), and is given by
\begin{align}
\label{eq:gdef}
\nonumber &g(n,e) = \frac{n^4}{32} \biggl \{ \biggl [ J_{n-2}(ne) - 2e J_{n-1}(ne) + \frac{2}{n} J_n(ne) \\
\nonumber &\quad + 2e J_{n+1}(ne) - J_{n+2}(ne) \biggl ]^2 + \left(1-e^2\right) \biggl [J_{n-2}(ne) \\
&\quad- 2 J_n (ne)+J_{n+2}(ne) \biggl ]^2 + \frac{4}{3n^2} J_n(ne)^2 \biggl \}.
\end{align}
Here, $J_k(x)$ refers to Bessel functions of the first kind. For a given eccentricity, the peak power is emitted at a harmonic where $g(n, e)$ has the maximum value. For ease of calculation, \citet{2003ApJ...598..419W} provided a fit to the peak harmonic as a function of eccentricity,
\begin{equation}
    n^{\mathrm{Wen}}_\mathrm{peak}(e) = 2 \dfrac{(1 +e)^{1.1954}}{(1 - e^2)^{3/2}} \qq{.}
\end{equation}
\citet{2021RNAAS...5..275H} noted that the above fit does not work well at low eccentricities, and provided an alternative fit to the peak harmonic,
\begin{equation}
    n^{\mathrm{Hamers}}_\mathrm{peak}(e) = 2 \qty(1 + \sum_{k=1}^{4} c_k e^k){(1 - e^2)^{-3/2}} \qq{.}
\end{equation}
Here, $c_1=-1.01678$, $c_2 = 5.57372$, $c_3 = -4.9271$, and $c_4 = 1.68506$.
Both these estimates are plotted in Figure~\ref{fig:peak_harmonic} along with the peak harmonic calculated numerically.

\subsection{Eccentricity extraction prescription of \citet{2003ApJ...598..419W}}\label{subsec:WenFormalism}
Since the value of eccentricity changes during the course of the inspiral, the value of eccentricity is often quoted at a specific reference frequency (or reference time). \citet{2003ApJ...598..419W} (henceforth referred to as W03) uses the peak harmonic's frequency as a reference for defining eccentricity, and this definition has been employed by many other studies (e.g.~\citealt{2017ApJ...840L..14S,2018PhRvD..97j3014S,2018PhRvD..98l3005R,2019ApJ...871...91Z,2021ApJ...921L..43Z, 2022arXiv221010055K, 2023MNRAS.526.4908C}). Given semimajor axis $a$ and total mass $M$, the average orbital frequency $f_\mathrm{orb}$ is 
\begin{equation}
    f_\mathrm{orb} = \dfrac{1}{2 \pi} \sqrt{\dfrac{GM}{a^3}} \qq{.}
\end{equation}

This means that the peak \ac{GW} frequency is,
\begin{equation}
\label{eq:fpeakGW_a_e}
f_\mathrm{peak, GW} =  \dfrac{(1 +e)^{1.1954}}{\pi(1 - e^2)^{3/2}}  \sqrt{\dfrac{GM}{a^3}}\qq{.}
\end{equation}
In astrophysical simulations, we are generally interested in obtaining eccentricity at a specified reference frequency, given an initial value of semi-major axis $a_0$ and eccentricity $e_0$. These initial values can be thought of as arising either from some fiducial distribution of eccentricities and separations, or from stopping conditions in dynamical simulations.
W03 uses the leading (Newtonian) order evolution~\citep{1964PhRv..136.1224P} of the semi-major axis and eccentricity to relate $(a_0, e_0)$ to $(a, e)$ at a different time,
\begin{widetext}
    
\begin{equation}
    \label{eq:a_of_e}
    \dfrac{a (1 - e^2)}{e^{12/19}} \qty(1 + \dfrac{121}{304} e^2)^{-870/2299} = \dfrac{a_0 (1 - e_0^2)}{e_0^{12/19}} \qty(1 + \dfrac{121}{304} e_0^2)^{-870/2299}\qq{.}
\end{equation}
\end{widetext}
For a given value of the reference frequency $f_{\rm peak, GW}$, the eccentricity can be obtained by solving Eq.~\eqref{eq:fpeakGW_a_e} and Eq.~\eqref{eq:a_of_e} simultaneously.

\subsection{Possible issues with the W03 prescription}
While the W03 procedure has been used in deriving eccentricities at a particular \ac{GW} frequency in many simulations, it has a couple of drawbacks:
\begin{enumerate}
\item The reference frequency is taken to be the peak frequency of the orbit. This is inconsistent with recommendations in the literature attempting to standardize the definition of eccentricity in waveforms~\citep{2022PhRvD.105d4035R,2023PhRvD.108j4007S}. Specfically, these works motivate using the orbit-averaged 22 (i.e. $\ell=2$, $m=2$) mode frequency as a consistent reference point, on account of its smooth and monotonic evolution over the course of the inspiral.
\item Since all equations are based on \citet{1964PhRv..136.1224P}, they are restricted to Newtonian (i.e. leading) order. This is not a good approximation of the evolution of the orbit, especially for systems close to merger. Specifically, the ordinary differential equations describing the evolution of eccentricities and the 22 mode frequencies have been calculated to several PN orders beyond the Newtonian order, also with the inclusion of spins. Furthermore, this equation breaks down at very high eccentricity.
\end{enumerate}
In the rest of our paper, we shall formulate a procedure that can be applied to any set of scattering experiments to extract eccentricities, and conduct sanity checks of our procedure. In order to accomplish this, we shall solve the problems listed above\footnote{Beyond the drawbacks mentioned here, we also note that, on account of being orbit-averaged, the Peters equations break down when the orbital time scale $\tau_{\rm orb}$ is much greater than the radiation reaction time scale $\tau_{\rm rr}$ (e.g. at large separations and high eccentricity; see for instance the discussion in \citealt{2023PhRvD.108l4055F} and references therein.). However, since end states of binaries from typical simulations follow $\tau_{\rm rr} \gg \tau_{\rm orb}$, we do not attempt to account for this issue. }.

\section{A consistent prescription for calculating eccentricity}\label{sec:standardized_eccentricity}

As shown in Figure~\ref{fig:peak_harmonic}, the peak harmonic of an eccentric inspiral takes higher values for higher eccentricities. For low values of eccentricity ($e \lesssim 0.3$), the peak harmonic is $n_{\rm peak} = 2$, and hence the 22-mode frequency is the same as frequency of the peak harmonic. 
However, the difference between the peak harmonic frequency and the 22-mode frequency increases very quickly, with $f_{\rm peak} = 10 f_{\rm orb}= 5 f_{22}$ at $e \approx 0.7$. This discrepancy means that the eccentricities derived from astrophysical simulations of different binary formation channels is defined at a different reference frequency as compared to the definition of reference frequency built into waveforms as per recommendations of \citet{2023PhRvD.108j4007S}. As is also evident, this discrepancy is worse for higher eccentricities, which are arguably more interesting from an astrophysical standpoint. 

To remedy this discrepancy, the definitions of eccentricity from the simulations and waveforms need to be placed on the same footing. Hence, instead of defining eccentricity at the frequency of the peak harmonic, we define eccentricity at the orbit-averaged 22 mode frequency.  Let $M=m_1+m_2$ be the total mass of the binary and $\eta = (m_1 m_2) / M^2 $ be the symmetric mass ratio, where $m_1$ and $m_2$ are the component binary masses. Given an initial value of semimajor axis $a_0$ and eccentricity $e_0$ for a particular binary from a simulation, we evolve  differential equations in $e$ and $v$ given by \ac{PN} formulae,
\begin{widetext}
\begin{align}
    \label{eq:dedt}
    \dv{e}{t} &= -\frac{e (121 e^2 +304) \eta  v^8}{15 M \epsilon ^5}\qty[1 + \alpha_2 v^2 + \alpha_3 v^3 + \alpha_4 v^4 + \order{v^5}]\\
    \label{eq:dvdt}
    \dv{v}{t} &= \frac{\left(37 e^4+292 e^2+96\right) \eta  v^9}{15 M \epsilon ^7}\qty[1 + \beta_2 v^2 + \beta_3 v^3 + \beta_4 v^4 + \order{v^5}] \\
    \label{eq:dfdt}
    f_{22} &= \dfrac{v^3}{\pi M}
\end{align}
\end{widetext}
where $\epsilon = \sqrt{1 - e^2}$, and $\alpha_i$, $\beta_i$ are \ac{PN} coefficients that depend on $e$ and $\eta$~\citep{1964PhRv..136.1224P, 1992MNRAS.254..146J, 1997CQGra..14.2357R,1997PhRvD..56.7708G,2009PhRvD..80l4018A}\footnote{See Appendix~\ref{app:PN-coeff} for explicit expressions of these.}. For the purposes of this work, we will  use equations accurate up to 2\ac{PN} order (i.e. terms up to $\order{v^4}$), and will neglect the effect of spin. We note that the full expressions are known up to 3\ac{PN} order~\citep{2009PhRvD..80l4018A} including the effect of spin~\citep{2023PhRvD.108j4016H}, but we find that restricting to 2\ac{PN} gives results accurate enough for our purposes\footnote{We could also equivalently use the approximate analytical expression generalizing Eq.~\eqref{eq:a_of_e} derived by \citet{2021PhRvD.104j4023T} instead of evolving the \ac{PN} differential equations, but we concentrate on the latter in this work.}.
Evolving the above equations in time starting from the initial conditions, we extract eccentricities at an appropriate reference frequency. This way of defining eccentricity makes it consistent with the one defined in waveforms, modulo systematics in modeling the \ac{GW} emission itself. \corrections{In the next subsection, we compare estimates derived from the prescription mentioned  above (which we denote by $e_t^{\rm 2 PN}$) to those derived from the W03 prescription (denoted as $e_{\rm W03}$) as used in the astrophysical simulations literature, and also investigate potential biases due to incomplete physics}.
\begin{figure*}[t!]
    \centering
    \includegraphics{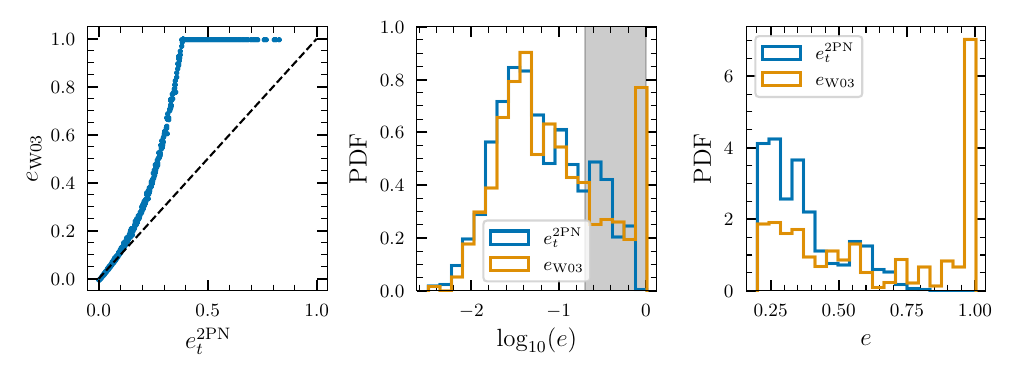}
    \caption{\corrections{Comparison of orbital eccentricities in \ac{CMC} simulations derived using the W03 prescription ($e_{\rm W03}$), to ones derived  using the prescription in this work ($e_t^{\rm 2 PN}$) at a reference frequency of $10\,{\rm Hz}$ set by each definition}. The left panel shows a scatter plot of the two quantities. We see that there is considerable disagreement on the derived values of eccentricity, especially at $e \gtrsim 0.2$. This can be understood as the effect of the peak frequencies being shifted to higher $n_\mathrm{peak}$. The middle panel shows the histogram  of (log) eccentricities for \ac{GW} captures in globular clusters that occur during strong, few-body resonating interactions, derived using the W03 prescription and the prescription described in this work. The histograms do not deviate significantly from each other at low values of eccentricity, but are significantly different at $e \gtrsim 0.2$ (shaded region). The right panel shows the histogram of eccentricities in this $e \gtrsim 0.2$ region, where the discrepancies are more evident. Notably, the peak at $e \approx 1$ is no longer present in the histogram; this peak is an artifact of the procedure used to extract eccentricity in astrophysical simulations. }
    \label{fig:comparison-of-ecc}
\end{figure*}

\subsection{Comparison with the W03 prescription}
\label{subsec:comparison-old-new}

As an illustration of our prescription, we consider binary mergers simulated with the Cluster Monte Carlo (\ac{CMC}) code~\citep{2022ApJS..258...22R}.  \ac{CMC} performs long-term evolution of a globular cluster using a H\'enon type Monte Carlo algorithm~\citep{henon1971a, henon1971b, joshi2000, joshi2001, fregeau2003, fregeau2007, chatterjee2010, chatterjee2013, pattabiraman2013, rodriguez2015} given a set of initial conditions for the cluster. While we will only consider the \ac{CMC} simulations as an example in this section, we again note that the W03 prescription has been in many other works (see e.g.~\citealt{2017ApJ...840L..14S,2018PhRvD..97j3014S,2018PhRvD..98l3005R,2019ApJ...871...91Z, 2022arXiv221010055K, 2023MNRAS.526.4908C}).

A catalog of \ac{CMC}  simulations on a grid of initial cluster masses,  metallicity, radii, and galactocentric distances is publicly available~\citep{cmc-paper, cmc-url}. This catalog also includes information about compact object mergers that take place in these clusters. Merging compact binaries are extracted from these simulations using the following procedure:
\begin{enumerate}
    \item During strong \ac{GW} encounters, the approximated $N$-body dynamics (including leading-order \ac{PN} corrections that account for orbital energy dissipation from \ac{GW} emission~\citep{2018PhRvL.120o1101R}) is evolved until two of the compact objects are in a bound state and reach a critical separation (typically $100 M$), at which point the binary is assumed to definitively lead to a \ac{GW}-driven merger and the properties of the binary are recorded.
    \item With this pair of eccentricity and semimajor axis, the eccentricity at a given reference frequency is extracted using the prescription of W03.
\end{enumerate}

Instead of relying on the W03 definition of eccentricity,  we apply the procedure outlined in Section~\ref{sec:standardized_eccentricity} to each binary extracted from the simulations. We then calculate eccentricities at a fixed 22-mode reference frequency of 10 Hz. 
In the left panel of  Figure~\ref{fig:comparison-of-ecc}, we show a scatter plot of  eccentricities derived from the two different definitions. Here, $e_{\rm W03}$ corresponds to the eccentricity as calculated by the W03 prescription, whereas $e_t^{\rm 2 PN}$ corresponds to the one calculated using the prescription presented in this work. 
At low eccentricities, the two definitions agree with each other. However, as one increases the eccentricity, the disagreement between the two increases quite rapidly. This is a direct result of the peak frequency getting shifted to higher harmonics as one increases eccentricity; for higher eccentricity, a given peak frequency $f_{\rm peak, GW}$ will correspond to a lower $f_{22}$ (since $ \frac{1}{2} n_{\rm peak}  f_{22} = f_{\rm peak, GW}$), hence describing the $e_t^{\rm 2PN}$ at a much lower reference frequency. This also means that the binary will lose more of its eccentricity by the time it reaches $f_{\rm 22}  = 10\,{\rm Hz} $.  As a consequence, $e_{t}^{\rm 2 PN}$ is always less than $e_{\rm W03}$. 

Notably, all the points that have $e_{\rm W03} \approx 1$ have $e_{t}^{\rm 2 PN}$ that range from $0.4-0.8$. We note that in many places in the literature, this $e \approx 1$ peak has been interpreted to be binaries that form ``in-band'' and merge quickly after getting captured; these are classically hyperbolic systems that are captured with a periapse frequency \textit{above} the reference frequency considered (e.g., 10 Hz) via orbital energy loss through \ac{GW} emission\footnote{We note that the $e \approx 1$ value for systems in astrophysical simulations is somewhat artificial. These represent systems that do not have a defined eccentricity at a particular reference frequency according to the prescription outlined in Section~\ref{subsec:WenFormalism} and thus correspond to $e \geq 1$ at a particular reference frequency.}. While that interpretation still remains qualitatively correct, our analysis shows that the value of eccentricity that should be supplied to waveforms in comparison studies is not $e \approx 1$ but rather is shifted to much lower values.

The middle panel of Figure~\ref{fig:comparison-of-ecc} shows histograms of $e_{\rm W03}$ and $e_{t}^{\rm 2 PN}$ calculated from the \ac{CMC} simulations, with appropriate weights applied to each sample based on cluster mass, metallicity, and detection probability\footnote{See e.g.~Appendix A of \citet{2021ApJ...921L..43Z} (and references therein) for a description of this weighting procedure.}. As would be expected from the preceding discussions, the histograms match for low values of eccentricity, and start deviating at $e \gtrsim 0.2$. Most strikingly, the peak at $e_{\rm W03} \approx 1 $ in the $e_{\rm W03}$ histogram is not present in the $e_t^{\rm 2 PN}$ histogram. These observations are more evident in the right panel of Figure~\ref{fig:comparison-of-ecc}, where we show the histograms of eccentricities for $e > 0.2$. Since the waveform changes significantly as a function of eccentricity, one would expect these discrepancies to impact any analysis that assumes consistency between the definitions of eccentricities in waveforms and astrophysical simulations. We will comment on this briefly in Section~\ref{subsec:detectability}.

\begin{figure}[!htbp]
    \centering
    \includegraphics{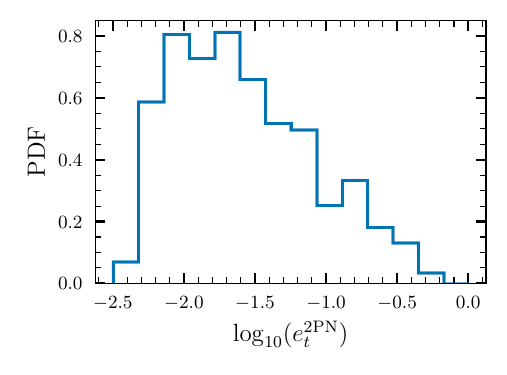}
    \caption{Histogram of eccentricities extracted at $Mf_{ 22} = 1000 \, M_\odot{\rm Hz}$. Defining the eccentricity at a fixed $M f_{22}$ has a few advantages: (i) $M f_{22}$ is independent of redshift,  (ii) it defines eccentricity consistently at a fixed number of cycles before merger, and (iii) it is easy to implement while estimating source parameters from \ac{BBH} signals. We recommend that all astrophysical simulations provide $e_t^{\rm 2 PN}$ estimates extracted at a fixed $Mf_{22}$ for easy comparison with estimates of eccentricity derived using GW waveform models.}
    \label{fig:e_Mf}

\end{figure}

In the preceding discussion, we have assumed that the reference frequency is defined in the source-frame of the binary as opposed to the detector-frame. We make this choice since the original calculation of eccentricity using the W03 prescription with the \ac{CMC} outputs made this choice. Strictly, this is incorrect and adds another significant level of inconsistency in the conventions for eccentricity. For completeness, we also show the eccentricity extracted at a reference frequency corresponding to a constant  $f_{\rm 22} M = 1000\,{\rm Hz}M_\odot$ in Figure~\ref{fig:e_Mf}, as $f_{\rm 22} M$ does not depend on redshift. At a $f_{\rm 22}$ corresponding to constant $f_{\rm 22} M$, all (black hole) binaries would be at the same epoch in their orbital evolution, i.e. they will all be at a fixed number of cycles before merger. This criterion is easy to implement while estimating parameters from \ac{BBH} signals, and also allows for a more direct comparison of eccentricities for systems with different masses.

It is natural to ask whether the discrepancy is primarily driven by the mismatch in conventions for the reference frequency, or due to only accounting for the leading order PN term in the W03 prescription. The results in the preceding paragraphs would lead us to believe that it is the former. We test this in another way: since we know $n_{\rm peak}(e)$, we can also adjust the eccentricities calculated via W03 as a post-processing step by assuming that the $e_{\rm W03}$ is defined at a reference frequency $f_{\rm 22} = 2 \times f_{\rm peak} / n_{\rm peak}(e) $, and evolve to a fixed reference (22-mode) frequency using PN equations (Eqs.~\eqref{eq:dedt}-\eqref{eq:dfdt}) but only up to the leading order. Let the eccentricity calculated using this modified version of the W03 procedure be $e_{\rm W03}^{f_{22}}$. We find that $e_{\rm W03}^{f_{22}} \approx e_{t}^{\rm 2 PN}$ for the binaries we consider\footnote{We exclude binaries with $e_{\rm W03} \approx 1$ from this calculation, due to their aforementioned artificial nature.}, with a $\sim 5\%$ scatter that could be attributed to non-inclusion of higher-order PN terms (see also Section~\ref{subsec:higher-order-pn-terms}). This further confirms that the discrepancy in the calculated eccentricities is primarily driven by the difference in conventions for the reference frequency.

\begin{figure}[!htbp]
    \centering
    \includegraphics{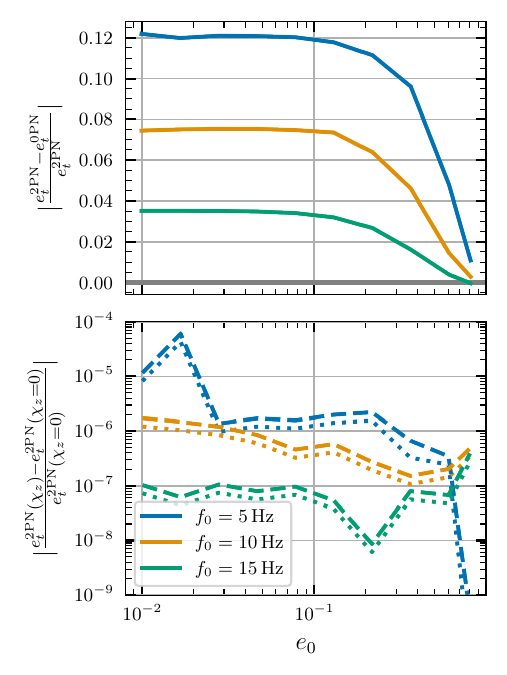}
    \caption{The top panel compares $e_t^{\rm 0PN}$ and $e_t^{\rm 2PN}$ computed at $f_{\mathrm{ref}} = 20 \, \rm Hz$ for \corrections{three  22-mode starting frequencies} $f_0=5,10, 15\,{\rm Hz}$ as a function of initial eccentricity $e_0$. The bottom panel makes a similar comparison between $e_t^{\rm 2PN}$ calculated assuming a value $\chi_{ z}=\chi_{1, z}=\chi_{ 2, z}$, and $\chi_z=0$. We choose two representative values of $\chi_z$: $\chi_z = 0.7$ (dotted lines), and $\chi_z = 0.99$ (dashed lines).}
    \label{fig:higher-PN-spin}
\end{figure}

\subsection{Effects of ignoring higher order PN terms and spin effects}
\label{subsec:higher-order-pn-terms}

We now systematically investigate if there is a relative advantage in using the 2\ac{PN} equations above just the leading order (0\ac{PN})  evolution equations. For an equal mass binary of total mass $M=50\,M_\odot$, we consider three different \corrections{22-mode starting frequencies} $f_0=5,10,15\, {\rm Hz}$, and calculate $e_t^{\rm 2 PN}$ and $e_t^{\rm 0 PN}$ at a reference frequency of 20 Hz for a range of initial eccentricities $e_0$. The fractional differences between $e_t^{\rm 2 PN}$ and $e_t^{\rm 0 PN}$ estimates are plotted in the top panel of Figure~\ref{fig:higher-PN-spin}, and show an increasing trend for lower values of $f_0$, with a discrepancy as high as $12\%$ for $f_0 = 5\, {\rm Hz}$. For a given $f_0$, the fractional differences are lower for higher $e_0$. In typical applications, one might want to calculate eccentricities at a given reference frequency for much lower values of $f_0$. While a $\sim10\%$ difference might not be relevant while comparing eccentricity estimates from data to simulations,  we recommend using $e_t^{\rm 2PN}$ estimates since the costs of evaluating $e_t^{\rm 0PN}$ and $e_t^{\rm 2PN}$ are similar (see Appendix~\ref{app:time-calculations}).

Our prescription in Section~\ref{sec:standardized_eccentricity} does not include the effects of spin. To see if this is a valid choice, we use the same mass, $f_0$, $f_{\rm ref}$, and $e_0$ values as in the previous paragraph and calculate $e_t^{\rm 2PN}$ with and without including spin terms in the evolution for two representative dimensionless (aligned) spin values of $\chi_{1,z} = \chi_{2,z} = \chi_{z} = [0.7,\,0.99]$. For the spin terms in the evolution, we use 2\ac{PN} accurate terms from \citet{2018PhRvD..98j4043K}. The fractional differences between  $e_t^{\rm 2PN}$ estimates with and without spin effects are shown in the lower panel of Figure~\ref{fig:higher-PN-spin}. We see that the discrepancies are at worst $\sim 10^{-5}$, shifting to even lower values when considering smaller fiducial spins or higher $f_0$. We hence conclude that the non-spinning approximation is sufficient.

\subsection{Comparison to estimates derived from \texttt{gw\_eccentricity} package}
\label{subsec:gw-eccentricity}

\begin{figure}[b!]
    \centering    \includegraphics{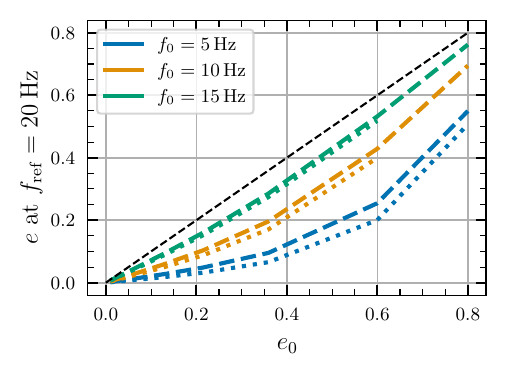}
    \caption{Comparison between the eccentricity definition given by \citet{2023PhRvD.108j4007S},  calculated using the waveform approximant \texttt{SEOBNRv4EHM} ($e_{\rm gw}^{\rm SEOB}$; dotted lines), and $e_t^{\rm 2PN}$ (dashed lines), as a function of the initial eccentricity, $e_0$, for three initial frequencies, $f_0=5,10,15\,{\rm Hz}$. We assume an equal mass non-spinning binary with total mass $M=50\,M_\odot$ for this purpose.}
    \label{fig:egw-vs-et}
\end{figure}

As mentioned earlier, different \ac{GW} waveform approximants have different conventions to define eccentricity. To solve this issue, \citet{2023PhRvD.108j4007S} proposed a standardized definition of eccentricity that can be applied to any waveform approximant, $e_{\rm gw}$, which is given by 
\begin{equation}
e_{\rm gw} = \cos{\left(\Psi/3\right)} - \sqrt{3}\sin{\left(\Psi/3\right)} \;,
\end{equation}
where 
\begin{eqnarray}
\Psi &=& \arctan{\left(\frac{1 - e_{\omega_{22}}^2}{2 e_{\omega_{22}}}\right)}\;, \\
e_{\omega_{22}}(t) &=& \frac{\sqrt{\omega_{22}^\mathrm{p}(t)} - \sqrt{\omega_{22}^\mathrm{a}(t)}}{\sqrt{\omega_{22}^\mathrm{p}(t)} + \sqrt{\omega_{22}^\mathrm{a}(t)}} \;,
\end{eqnarray}
and $\omega_{22}^{\rm p,a}(t)$ are the 22-mode frequencies through the pericenters and apocenters, respectively. This standardized definition is proposed in order to extract eccentricity directly from observables, in this case the gravitational waveform. More details on the motivations for this definition and the various methods of calculating $\omega_{22}^{\rm p, a} (t)$ can be found in \citet{2023PhRvD.108j4007S}. We also briefly investigate the differences between $e_{\rm gw}$ derived from different waveform approximants in Appendix~\ref{app:gw-eccentricity-differences}.

While $e_{\rm gw}$ does asymptote to the \ac{PN} definition of eccentricity in the \corrections{small eccentricity limit as well as the large separation limit}, these two definitions may differ in general. For instance, waveforms could include more accurate dynamics at high eccentricity and closer to merger as compared to our 2\ac{PN} prescription. Instead of the 2\ac{PN} accurate equations, we could hence imagine replacing the prescription in Section~\ref{sec:standardized_eccentricity} by generating waveforms at the specified initial epochs, and let $\texttt{gw\_eccentricity}$ calculate the eccentricities at the required reference frequency. 

However, since calculating $e_{\rm gw}$ involves computing the waveform starting at a specified initial frequency/time, the calculation would be on the order of a hundred times slower (see Appendix~\ref{app:time-calculations}) than computing the \ac{PN} eccentricity. The cost of generating waveforms also increases with decreasing starting frequency. Since astrophysical simulations require calculating eccentricites of a large number of binaries, 
calculating the \citet{2023PhRvD.108j4007S} definition of eccentricity would be computationally inefficient. Therefore, we now compare the differences between the two eccentricity definitions, $e_{\rm gw}$ and $e_t^{\rm 2PN}$, and quantify the reliability of $e_t^{\rm 2PN}$ eccentricity estimates from astrophysical simulations. For the investigations that follow, we will use the \texttt{SEOBNRv4EHM}~\citep{2022PhRvD.105d4035R} waveform approximant for the calculation of $e_{\rm gw}$. Using the same mass, $f_0$, and $e_0$ values as in Section~\ref{subsec:higher-order-pn-terms}, we calculate $e_t^{\rm2 PN}$ and $e_{\rm gw}^{\rm SEOB}$  at a reference frequency of 20 Hz. Figure \ref{fig:egw-vs-et} illustrates the differences between these. While small (a few tens of a percent at worst), the differences do accumulate as the binary evolves for a longer period of time with $f_0$ values closer to the reference frequency giving more agreement. Since the cost of computing $e_{\rm gw}$ is a factor of $\sim 100$ greater as compared to calculating $e_t^{\rm 2PN}$ (see Appendix~\ref{app:time-calculations}), we conclude that $e_t^{\rm 2PN}$ is sufficient, especially in regimes where \ac{PN} theory is valid.

\subsection{Implications for existing work}
\label{subsec:detectability}
A number of works have ascertained the detectability of eccentric mergers within the LVK band as well as in other  detector bands, using data from astrophysical simulations such as \ac{CMC}. As we have shown in the previous sections, the waveform eccentricities are not in agreement with the eccentricities given by simulations. Hence, we would expect these detectability estimates to change when defining eccentricity consistently in the simulations. For instance, \citet{2021ApJ...921L..43Z} found that $\approx${60\%} of potentially detectable eccentric sources with $e \gtrsim 0.2$ would be missed due to non-inclusion of eccentric templates in search pipelines. Using the same dataset as \citet{2021ApJ...921L..43Z} and changing the eccentricities from the $e_{\rm W03}$ definition to the $e_t^{\rm 2 PN}$ definition, we find that only 30\% of potentially detectable eccentric sources with $e \gtrsim 0.2$ are missed---a discrepancy by a factor of $\approx 2$. The decrease in the number of missed signals is due to the shifting of the histogram of eccentricities towards lower values, which are better approximated by non-eccentric templates and hence less likely to be selected against. As another illustration, we calculate limits on the cluster branching fraction $\beta_c$ i.e the fraction of events in the total population coming from clusters using the same method and dataset as used in \citet{2021ApJ...910..152Z}. We find that assuming $N_{\rm ecc} = 0$ eccentric detections in GWTC-3, we would have inferred $\beta_c < 86\%$ ($90\%$ credible level) using the $e_{\rm W03}$ estimates,  and infer $\beta_c < 63\%$ using the $e_t^{\rm 2 PN}$ estimates. On the other hand, assuming $N_{\rm ecc} = 1$ gives $\beta_c > 10\%$ using the $e_{\rm W03}$ estimates and $\beta_c > 8\%$ using the $e_t^{\rm 2 PN}$ estimates.

Similarly, any study that uses the  detected eccentricities (or lack thereof) to place constraints on the overall merger rate in eccentric systems in astrophysical environments  will also be affected by these discrepancies. 
For instance, \citet{2023arXiv230803822T} use the non-detection of orbital eccentricity from a waveform-independent search to place an upper limit of  $0.34\,{\rm Gpc^{-3}\,{yr}^{-1}}$ on the merger rate of binaries in dense stellar clusters, assuming a population model from \ac{CMC} simulations. This population model is taken from \citet{2021ApJ...921L..43Z} with eccentricities defined using the W03 prescription, with additional conditioning on the mass ($M > 70$) and eccentricity  ($e < 0.3$) parameter ranges of their search. From Figure~\ref{fig:comparison-of-ecc}, we see that the relative number of mergers having $e < 0.3$ increases when accounting for the prescription in this work. This means that an upper limit derived using the $e_t^{\rm 2 PN}$ estimates from this work would be lower than the ones quoted in \citet{2023arXiv230803822T}. 

We also note that \citet{2022ApJ...940..171R}, which uses signatures of non-zero eccentricity in four \ac{GW} events to constrain the fraction of \ac{BBH} mergers formed dynamically,  does account  for the disparate conventions for eccentricity while making their comparisons. They do so in an approximate fashion, similar to the calculation of $e_{\rm W03}^{f_{22}}$ described in Section~\ref{subsec:comparison-old-new}. We do not expect conclusions reported there to change significantly.

\section{Conclusion}\label{sec:conclusions}
Orbital eccentricity in a compact binary is believed to be a telltale sign of its formation history. Consequently, there has been progress on building waveform models for eccentric sources as well as on simulating expectations from astrophysical formation channels. 
In this work, we note that the conventions used to define eccentricity in the \ac{GW} waveform and astrophysical simulation communities differ from each other. The major difference in these conventions lies in the choice of the reference frequency at which eccentricity is defined---\ac{GW} waveforms typically define eccentricity at a fixed orbit averaged 22-mode \ac{GW} frequency, while astrophysical simulations define eccentricity at a fixed peak \ac{GW} frequency following the W03 prescription (described in Section~\ref{subsec:WenFormalism}). Further, we delineate an eccentricity calculation procedure that can be applied to data from astrophysical simulations to extract eccentricities consistent with conventions laid down by the \ac{GW} waveform community. This procedure involves evolution of ordinary differential equations describing the time-evolution of eccentricity and frequency obtained from \ac{PN} expansions, and extracting eccentricity consistently at the orbit-averaged 22 mode frequency.

Using compact binaries from the publicly-available \ac{CMC} catalog as an illustration, we find that there is a discrepancy between eccentricity estimates derived using the W03 prescription and the one described in this work. The discrepancy is small for $e \lesssim 0.1$, but is large for higher values of eccentricity. Notably, all binaries with $e_{\rm W03} \approx 1$ (thought to be \ac{GW} captures happening in the LVK band) have $e_t^{\rm 2 PN}$ between $0.4$--$0.8$. We investigate the effect that adding 2\ac{PN} corrections to the leading order evolution has on derived eccentricities, and also the effects of not including spins in our prescription. We also compare our $e_t^{\rm 2PN}$ estimates to $e_{\rm gw}$ estimates derived from the \texttt{gw\_eccentricity} package using the \texttt{SEOBNRv4EHM} waveform. \corrections{We find that the $e_t^{\rm 2 PN}$ estimates should be sufficiently accurate in extracting eccentricities, and also computationally efficient to implement as a post-processing step in simulations. However, as outlined in \citet{2023PhRvD.108j4007S}, \texttt{gw\_eccentricity} is the most accurate way of extracting eccentricity and should be used when high-precision eccentricity estimates are necessary. }

The different conventions for defining eccentricity currently employed in the literature mean that one should be careful in comparing eccentricity estimates/upper limits derived from real data to astrophysical simulations, creating mock populations of eccentric sources based on astrophysical expectations, and other investigations that pair predictions from astrophysical models with GW data analysis. As an illustration, we show that the detectability estimates of eccentric sources and branching fractions from \citet{2021ApJ...921L..43Z} will change when ensuring a consistent definition of eccentricity.
{We recommend that catalogs created from astrophysical simulations include $e_t^{\rm 2PN}$ estimates (extracted at a constant $Mf_{22} $) for clarity and ease of direct comparison to \ac{GW} observations.}

\section*{Acknowledgements}
We are grateful to Isobel Romero-Shaw for a thorough reading of our draft and useful suggestions. We thank Kaushik Paul for insighful inputs regarding \ac{PN} equations and sharing code for the same, and Antoni Ramos-Buades for providing access to the \texttt{SEOBNRv4EHM} approximant. We also thank Vijay Varma, Prayush Kumar, Bala Iyer, and Akash Maurya for useful discussions. \corrections{Example python scripts~\citep{vijaykumar_2024_10974975}  and data products~\citep{vijaykumar_2024_10633005} are available on Zenodo}.

AV is supported by the Natural Sciences and Engineering Research Council of Canada (NSERC) (funding reference number 568580), and the Department of Atomic Energy, Government of India, under Project No. RTI4001.  
AV also acknowledges support by a Fulbright Program grant under the Fulbright-Nehru Doctoral Research Fellowship, sponsored by the Bureau of Educational and Cultural Affairs of the United States Department of State and administered by the Institute of International Education and the United States-India Educational Foundation. AGH is supported by NSF grants AST-2006645 and PHY2110507. 
AGH gratefully acknowledges the ARCS Foundation Scholar Award through the ARCS Foundation, Illinois Chapter with support from the Brinson Foundation.
MZ gratefully acknowledges funding from the Brinson Foundation in support of astrophysics research at the Adler Planetarium.

\software{numpy~\citep{2020Natur.585..357H}, scipy~\citep{2020NatMe..17..261V}, matplotlib~\citep{2007CSE.....9...90H}, \corrections{astropy~\citep{2013A&A...558A..33A,2018AJ....156..123A,2022ApJ...935..167A}}, jupyter~\citep{2016ppap.book...87K}, pandas~\citep{mckinney-proc-scipy-2010}, seaborn~\citep{2021JOSS....6.3021W}, and lalsuite~\citep{lalsuite}.}

\newpage

\appendix 
\label{sec:appendix}
\onecolumngrid

\section{\ac{PN} formulae for eccentricity and velocity evolution}
\label{app:PN-coeff}
The \ac{PN} formulae (accurate up to 2\ac{PN}) for the evolution of eccentricity and velocity  are given below
\begin{align}
    \dv{e}{t} &= -\frac{e (121 e^2 +304) \eta  v^8}{15 M \epsilon ^5}\qty[1 + \alpha_2 v^2 + \alpha_3 v^3 + \alpha_4 v^4 + \order{v^5}]\\
    \dv{v}{t} &= \frac{\left(37 e^4+292 e^2+96\right) \eta  v^9}{15 M \epsilon ^7}\qty[1 + \beta_2 v^2 + \beta_3 v^3 + \beta_4 v^4 + \order{v^5}] \\
    f_{22} &= \dfrac{v^3}{\pi M}
\end{align}
where
\begin{align}
    \alpha_2 &= \frac{125361 e^4+718008 e^2-28 \left(3328 e^4+23259 e^2+8168\right) \eta -67608}{168 (121 e+304) \epsilon ^2}\\
    \alpha_3 &= \frac{1970 \pi  v^3 \epsilon ^5}{121 e+304} \phi_e^{\rm rad} \\
    \alpha_4 &= \frac{1}{2016 (121 e+304) \epsilon ^4} [3 e^6 \left(919520 \eta ^2-1448284 \eta +1262181\right) \nonumber \\
    &\mathrel{\phantom{=}} +e^4 \left(42810096 \eta ^2-78343602 \eta -569520 (2 \eta -5) \epsilon +46566110\right) \nonumber \\
    &\mathrel{\phantom{=}}-12 e^2 (3495771 \eta +584892 (2 \eta -5) \epsilon -945290) \nonumber \\
    &\mathrel{\phantom{=}}-16 \left(-284256 \eta ^2-801495 \eta +168336 (2 \eta -5) \epsilon +952397\right)] 
\end{align}
and
\begin{align}
    \beta_2 &= \dfrac{e^6 (11717-8288 \eta )+e^4 (171038-141708 \eta )-120 e^2 (1330 \eta -731)+16 (743-924 \eta )}{56 \left(37 e^4+292 e^2+96\right) \epsilon ^2}\\
    \beta_3 &= \frac{768 \pi  v^3 \epsilon ^7}{74 e^4+584 e^2+192} \phi_e\\
    \beta_4 &= -\frac{1}{6048 \left(37 e^4+292 e^2+96\right) \epsilon ^4} [-3 e^8 \left(654752 \eta ^2-1086660 \eta +1174371\right) \nonumber \\ 
     &\mathrel{\phantom{=}} +6 e^6 \left(-10804808 \eta ^2+20518071 \eta +88200 (2 \eta -5) \epsilon -13904067\right) \nonumber \\
     &\mathrel{\phantom{=}}+12 e^4 \left(-13875505 \eta ^2+17267022 \eta +1105272 (2 \eta -5) \epsilon -65314\right) \nonumber \\ 
     &\mathrel{\phantom{=}}+16 e^2 \left(-3830127 \eta ^2-966546 \eta +806652 (2 \eta -5) \epsilon +5802910\right) \nonumber \\
     &\mathrel{\phantom{=}}+32 \left(-59472 \eta ^2-141093 \eta +9072 (2 \eta -5) \epsilon +11257\right)] \qq{.}
\end{align}
Here, $\epsilon = \sqrt{1 - e^2}$ and the ``enhancement factors'' $\phi_e$,  $\tilde{\phi}_e$, and $\phi_e^{\rm rad}$ are given by

\begin{align}
    \phi_e &= \frac{\frac{428340 e^{12}}{9958749469}-\frac{5034498 e^{10}}{7491716851}+\frac{9293260 e^8}{3542508891}+\frac{48176523 e^6}{177473701}+\frac{157473274 e^4}{30734301}+\frac{18970894028 e^2}{2649026657}+1}{\epsilon ^{10}} \\
    \tilde{\phi}_e &= \frac{-\frac{328675 e^{10}}{8674876481}-\frac{4679700 e^8}{6316712563}-\frac{2640201 e^6}{993226448}+\frac{37570495 e^4}{98143337}+\frac{413137256 e^2}{136292703}+1}{\epsilon ^9} \\
    \phi_e^{\rm rad} &= \frac{192 \epsilon (\epsilon \phi_e- \tilde{\phi}_e)}{985 e^2} \qq{.}
\end{align}

\section{Variation in \texttt{gw\_eccentricity} estimates calculated from different waveform approximants}
\label{app:gw-eccentricity-differences}

\begin{figure}[!tbp]
\centering
    \includegraphics{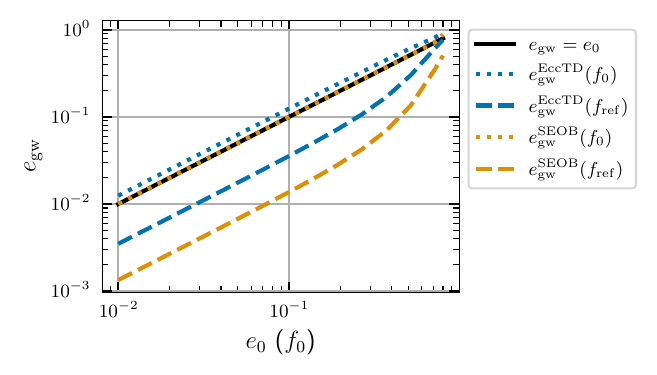}
    \caption{Comparison of internal definitions of eccentricity for different waveform models as a function of the initial input eccentricity, $e_0$ with a fiducial input frequency of 5 Hz. Initial eccentricities using the standardized definition, $e_{\rm gw}(f_0)$ (dotted lines), should be consistent with the input eccentricity $e_0$ (black line). \texttt{EccentricTD} (blue) does not agree for any value of $e_0$, while \texttt{SEOBNRv4EHM} (yellow) does agree for all $e_0$ values considered. We also see that the final eccentricity reported for the given reference frequency $f_{\mathrm{ref}}$ between the two waveforms (blue and yellow dashed lines) do not agree.}
    \label{fig:waveform-comparison}
\end{figure}

The prescription of calculating eccentricity $e_{\rm gw}$, proposed by \citet{2023PhRvD.108j4007S} and implemented in the \texttt{gw\_eccentricity} python package only uses the \ac{GW} waveform as seen by the detector as input, as described in Section~\ref{subsec:gw-eccentricity}. 
This definition is beneficial since eccentricity is now linked to a direct observable. However, different waveforms will, in general, yield different values of $e_{\rm gw}$ for the same binary. 
This could be due to a variety of reasons: differences in the modeling of the dynamics, different internal definitions of eccentricity in the model, or even choices made while implementing the approximant in some codebase such as \texttt{LALSimulation}~\citep{lalsuite}.
Figure \ref{fig:waveform-comparison} illustrates this point by comparing $e_{\rm gw}$ derived from two waveforms, \texttt{EccentricTD}~\citep{2016PhRvD..93f4031T} and \texttt{SEOBNRv4EHM}~\citep{2022PhRvD.105d4035R}. For this illustration, we assume an equal mass binary with a total mass of $M=50\, M_\odot$. We vary the initial eccentricity $e_0$ (defined at initial frequency $f_0 = 5\, {\rm Hz}$) supplied to the model between $10^{-2}$ and $0.8$, and extract $e_{\rm gw}$ from both \texttt{EccentricTD} $\qty(e_{\rm gw}^{\rm EccTD})$ and \texttt{SEOBNRv4EHM} $\qty(e_{\rm gw}^{\rm SEOB})$, at both the initial frequency $f_0$ and also at $f_{\rm ref} = 20\,{\rm Hz}$. We see that $ e_{\rm gw}^{\rm SEOB}(f_0)$ matches all input values $e_0$, whereas $ e_{\rm gw}^{\rm EccTD}(f_0)$ differs from $e_0$ by a few percent. This is in agreement with investigations performed in \citet{2023PhRvD.108j4007S}, where they found that the  $e_{\rm gw}^{\rm  SEOB} \qty(f_0) = e_0$  trend is maintained for all $e_0 > 10^{-5}$, whereas $e_{\rm gw}^{\rm EccTD}$ deviates from the $e_{\rm gw}^{\rm  EccTD} \qty(f_0) = e_0$  trend. We also see that there is a large discrepancy (factor of a few) between $e_{\rm gw}^{\rm EccTD}  (f_{\rm ref})$ and $ e_{\rm gw}^{\rm SEOB} (f_{\rm ref}) $, strengthening the evidence for differences in the ways each of these waveform approximants define eccentricity. We use the $\texttt{SEOBNRv4EHM}$ approximant for our comparisons in Section~\ref{subsec:gw-eccentricity}.

\section{Computational cost of different eccentricity prescriptions}
\label{app:time-calculations}
\begin{figure*}[!tbp]
    \centering
    \includegraphics{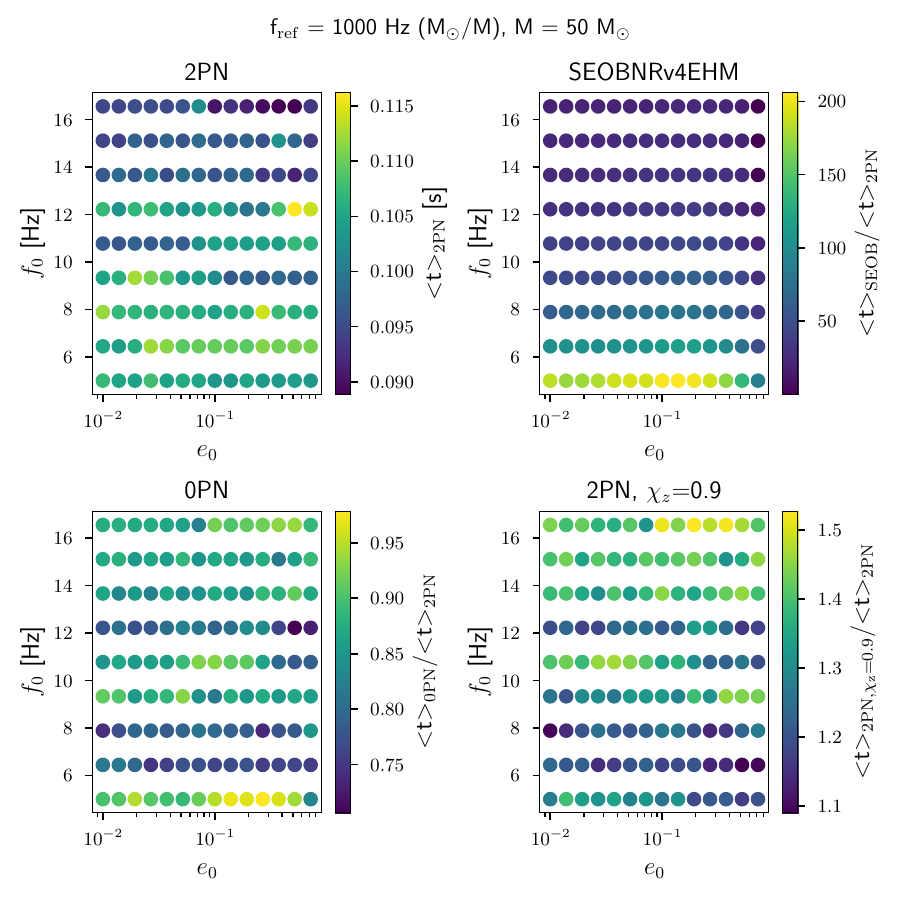}
    \caption{Average time to compute the 2\ac{PN} eccentricity (top left) and the relative time to compute the eccentricities using the \texttt{SEOBNRv4EHM} waveform (top right), the 0\ac{PN} eccentricity (bottom left), and the 2\ac{PN} eccentricity for spinning binaries (bottom right). The grid covers various input eccentricity, $e_0$, and frequency, $f_0$, values. Calculating eccentricities using waveforms takes $\sim 100\times$ longer than calculating eccentricities using our \ac{PN} approximation.}
    \label{fig:timeits}
\end{figure*}

Astrophysical simulations keep track of many evolving binary systems, where it is often computationally infeasible to compute waveforms for all systems. In Figure \ref{fig:timeits}, we demonstrate the speed-up in eccentricity calculation at the reference frequency, $f_{\rm ref}$, by computing the \ac{PN} eccentricities, $e_t^{k {\rm PN}}$, as opposed to the standardized waveform definition of eccentricity, $e_{\rm gw}$. The top left of Figure \ref{fig:timeits} demonstrates the average time to compute the 2\ac{PN} eccentricity at various initial eccentricities, $e_0$, and frequencies, $f_0$. The top right plot illustrates that computing the waveform eccentricity, $e_{\rm gw}$ using the \texttt{SEOBNRv4EHM} waveform model takes on the order of a hundred times longer on average. The bottom plots in Figure \ref{fig:timeits} demonstrate that using the 0\ac{PN} eccentricity has a comparable cost to using the 2\ac{PN} eccentricity, whereas including spin effects in the PN equations is $\sim$1.5 times slower. 

\twocolumngrid


\bibliography{standard_eccentricity}{}

\end{document}

%% file: commands.tex

\newcommand{\num}[1]{{\color{red} #1}}
\newcommand{\prelim}[1]{{\color[rgb]{0.75,0.0,0.0}#1}}
\definecolor{chmagenta}{rgb}{0.54, 0.17, 0.88}
\newcommand{\mjz}[1]{\textbf{\color{chmagenta} MJZ: #1}}

\newcommand{\changed}[1]{\textbf{#1}}
\newcommand{\changedmath}[1]{\mathbf{#1}}

\def\Mc{\ensuremath{\mathcal{M}_\mathrm{c}}\xspace}
\def\chieff{\ensuremath{\chi_\mathrm{eff}}\xspace}
\def\q{\ensuremath{q}\xspace}
\def\z{\ensuremath{z}\xspace}

\def\Msun{\ensuremath{\mathit{M_\odot}}\xspace}
\def\Rsun{\ensuremath{\mathit{R_\odot}}\xspace}
\def\Zsun{\ensuremath{\mathit{Z_\odot}}\xspace}

%% file: values.tex


%% file: acronyms.tex

\acrodef{GW}{gravitational wave}
\acrodef{BH}{black hole}
\acrodef{BBH}{binary black hole}
\acrodef{NS}{neutron star}
\acrodef{BNS}{binary neutron star}
\acrodef{EM}{electromagnetic}
\acrodef{PN}{post-Newtonian}
\acrodef{CBC}{compact binary coalescence}
\acrodef{CMC}{Cluster Monte Carlo}

\acrodef{LIGO}{Laser Interferometer Gravitational-wave Observatory}
\acrodef{LVK}{LIGO--Virgo--KAGRA}
\acrodef{LVC}{LIGO Scientific and Virgo Collaboration}
\acrodef{O1}{first observing run}
\acrodef{O2}{second observing run}
\acrodef{O3}{third observing run}
\acrodef{O3a}{first half of the third observing run}